\documentclass[aip,jmp,reprint,onecolumn,showpacs,superscriptaddress,amsfonts,amssymb,amsmath]{revtex4-1}
\usepackage[american]{babel}
\usepackage[latin1]{inputenc}
\usepackage[T1]{fontenc}

\usepackage{bm}
\usepackage{amssymb}
\usepackage{amsmath}

\newcommand{\Rbb}{\mathbb R}
\newcommand{\Cbb}{\mathbb C}

\newcommand{\be}{\begin{equation}}
\newcommand{\ee}{\end{equation}}

\newcommand{\ud}{\mathrm{d}}
\newcommand{\findiff}{\Delta}

\renewcommand{\Re}{\operatorname{Re}}

\begin{document}

\preprint{Niels Bohr Institute} 

\title{New Bessel Identities from Laguerre Polynomials} 

\author{Asger C. Ipsen}
 \affiliation{Niels Bohr International Academy and Discovery Center,
Niels Bohr Institute,
Blegdamsvej 17, DK-2100 Copenhagen, Denmark}


\begin{abstract}
For large order, Laguerre polynomials can be approximated by Bessel functions
near the origin. This can be used to turn many Laguerre identities into corresponding
identities for Bessel functions. We will illustrate this idea with a number of examples.
In particular, we will derive a generalization of a identity due to Sonine, which
appears to be new.
\end{abstract}


\maketitle
Bessel functions are ubiquitous within mathematical physics, and there
are a large number of sum and integral identities known for them. The 
same is true for Laguerre polynomials. Since a particular limit of the
Laguerre polynomials is given by Bessel functions, it is natural 
to consider whether some of the Laguerre identities could get a new life
as Bessel identities.

What  inspired this note, was an identity that showed up in the
solution of a random matrix model\cite{AD,AI}.
Many random matrix models are solvable, for finite matrix dimensions $N$,
in terms of orthogonal polynomials\cite{Meh04}.
In general the interesting case is when $N$ goes to infinity. 
In the case of Gaussian ensembles, the solution is in terms of Laguerre 
(or Hermite) polynomials. If one is interested in the spectrum near the origin,
one is then naturally lead to the kind of calculations we consider here.

The well known property of Laguerre polynomials we shall make use of, is that, 
for $\alpha$ real and fixed,
\be
  \lim_{N\to\infty}N^{-\alpha}L_N^\alpha\left(\frac {x^2}{4N}\right) 
    = 2^\alpha x^{-\alpha}J_\alpha(x),
  \label{eq:L-assymp-plain} 
\ee
uniformly for $x$ on some bounded region of $\Cbb$ (eg. Theorem 8.1.3 of ref.
\onlinecite{Sze75}). Using this together with standard asymptotic estimates like
Stirling's approximation, the passage from Laguerre identities to identities for
Bessel functions is mostly straightforward. This, of course, requires that  
the occurrences of the Laguerre polynomials are such that one can approach 
asymptotic region.

A curious phenomenon can occur if Laguerre polynomials with negative 
are present. When converting sums over such polynomials to integrals, it can
be necessary to add so-called \emph{anomalous} terms (as noted in ref. \onlinecite{AD}).
We define Laguerre polynomials at negative index by continuation of 
the usual formula
\be
  L_m^\alpha(x) := \sum_{j=0}^m \frac{(-)^j}{j!}
    {m+\alpha \choose m-j}x^j
    = \sum_{j=0}^m \frac{(-)^j}{j!(m-j)!}
    (\alpha+n)(\alpha+n-1)\cdots(\alpha+j+1)x^j.
  \label{eq:L-cont}
\ee
For general $\alpha$, clearly $L_m^\alpha(0)\neq0$.
But, from the last expression, we see that the lowest order term of 
$L_m^{-\nu}(x)$, with $\nu$ a positive integer, is $x^\nu$ when $m\geq\nu$.
This already hints at the fact that something special happens at the 
negative integers. We will also see that
this is connected with the behavior of Bessel functions near the origin.

The idea of using the fundamental relation between Laguerre polynomials and 
Bessel functions is very general, and appears to be relatively unexplored.
We seek to illustrate this method, by providing a number of examples, in the
hope, and with the expectation that, others will find it useful.
Some of the identities, like \eqref{eq:g-sonine}, we prove rigorously, while
for others, like \eqref{eq:order-sum-ident}, a complete proof is beyond the 
scope of this paper.
We stress that we have checked all the identities numerically to high
precision.

Let us now consider some concrete examples.
In the following, it will be useful to have \eqref{eq:L-assymp-plain} in a 
slightly different form. For fixed $\alpha\in\Rbb$ and $x\in\Cbb$, we have
\be
  \lim_{rN\to\infty}N^{-\alpha}L_{rN}^\alpha\left(\frac {x^2}{4N}\right) 
    = 2^\alpha r^{\alpha/2} x^{-\alpha}J_\alpha(\sqrt r x),
  \label{eq:L-assymp}
\ee
which holds uniformly
for $r$ on a bounded interval 
of the positive real axis (excluding the origin).

\section{Sonine like identities}
In this section we will derive two identities giving closed expressions
for certain integrals of products of two Bessel functions. The first one
generalizes a formula due to Sonine.

The starting point is the known sum formula for Laguerre polynomials
(eq. 22.12.6 in ref. \onlinecite{AS})
\be
  \sum_{m=0}^N L_m^\alpha(X)L_{N-m}^\beta(Y)
    = L_N^{\alpha+\beta+1}(X+Y),
  \label{eq:laguerre-sum-ident}
\ee
with the abbreviations
\[
  X := \frac{x^2}{4N}\quad\text{and}\quad Y := \frac{y^2}{4N},
\]
which we will use from now on.
Naturally this holds for arbitrary complex $x,y,\alpha,\beta$, but let us for the
moment restrict to $\alpha,\beta>-1$. We can now let $N\to\infty$
on both sides, which will replace the polynomials by Bessel functions.
The intuition is that, using
\eqref{eq:L-assymp} we replace the polynomials with Bessel function, 
and that the LHS turns into an integral over $r$. 

This intuition turns out to be too naive in some cases, 
so let us work out the details.
Fix $x,y\neq 0$, and $\epsilon > 0$, and set $r=m/N$. First we choose $M$ such that 
\[
  \bigl|N^{-\alpha-\beta} L_{m}^\alpha(X)L_{N-m}^\beta(Y)
    -2^{\alpha+\beta} r^{\alpha/2}(1-r)^{\beta/2} x^{-\alpha} y^{-\beta}
    J_\alpha(\sqrt r x)J_\beta(\sqrt{1-r}y)\bigr| 
      \leq \epsilon,
\]
when $M\leq m \leq N-M$. 
It then follows that 
\be
  \left|N^{-\alpha-\beta-1}\sum_{m=M}^{N-M} 
      L_m^\alpha(X)L_{N-m}^\beta(Y)
    - 2^{\alpha+\beta}x^{-\alpha} y^{-\beta}N^{-1}\sum_{m=M}^{N-M}
      r^{\alpha/2}(1-r)^{\beta/2}J_\alpha(\sqrt r x)J_\beta(\sqrt{1-r}y)\right| \leq \epsilon.
  \label{eq:est1}
\ee
Now choose $N$ such that 
\be
  \left|N^{-1}\sum_{m=M}^{N-M}
      r^{\alpha/2}(1-r)^{\beta/2}J_\alpha(\sqrt r x)J_\beta(\sqrt{1-r}y)
    -\int_0^1\ud r\,
      r^{\alpha/2}(1-r)^{\beta/2}J_\alpha(\sqrt r x)J_\beta(\sqrt{1-r}y)\right| \leq
      2^{-\alpha-\beta}x^\alpha y^\beta\epsilon,
  \label{eq:est2}
\ee
and 
\be
  \left| \sum_{m=0}^{M-1}\left[L_m^\alpha(X)L_{N-m}^\beta(Y)
  +L_{N-m}^\alpha(X)L_{m}^\beta(Y)\right]\right| 
  \leq N^{\alpha+\beta+1}\epsilon.
  \label{eq:est3}
\ee
To see that the last condition is possible to satisfy because, note that e.g. the
first term in the sum grows as $N^\beta$, which is dominated by $N^{\alpha+\beta+1}$,
since $\alpha+1>0$.

Combining \eqref{eq:est1},\eqref{eq:est2} and \eqref{eq:est3} we conclude that
\[
  \lim_{N\to\infty} N^{-\alpha-\beta-1}
    \sum_{m=0}^N L_m^\alpha(X)L_{N-m}^\beta(Y)
  = 2^{\alpha+\beta}x^{-\alpha} y^{-\beta}\int_0^1\ud r\,
      r^{\alpha/2}(1-r)^{\beta/2}J_\alpha(\sqrt r x)J_\beta(\sqrt{1-r}y),
\]
since $\epsilon$ was arbitrary. Taking the corresponding limit of the RHS of 
\eqref{eq:laguerre-sum-ident}, we obtain the identity
\be
  x^{-\alpha}y^{-\beta}\int_0^1\ud r\,
      r^{\alpha/2}(1-r)^{\beta/2}J_\alpha(\sqrt r x)J_\beta(\sqrt{1-r}y)
  = 2(x^2+y^2)^{-(\alpha+\beta+1)/2}J_{\alpha+\beta+1}(\sqrt{x^2+y^2}).
  \label{eq:g-sonine-sans-anoms}
\ee
By complex continuation\footnote{Note that $x^{-\alpha}J_\alpha(x)$ is analytic 
the entire complex plane.}, this holds for arbitrary $\alpha,\beta$ with 
$\Re\alpha,\Re\beta > -1$, which ensures convergence of the integral.
This formula is known as Sonine's second integral\cite{Son}, and a proof
based on expanding the Bessel functions using an integral representation
can be found in ref. \onlinecite{Wat22}. 
Later we will sketch yet another derivation from an Laguerre integral identity.

It is possible to generalize \eqref{eq:g-sonine-sans-anoms}. We note that,
due to the relation $J_{-n}(x) = (-)^nJ_n(x)$, the integral of
\eqref{eq:g-sonine-sans-anoms} is also convergent when $\alpha$ and/or
$\beta$ are negative integers. However, the identity does not hold as it
stands, because we can no longer satisfy \eqref{eq:est3}. Let us for
definiteness consider 
the case where $\alpha$ is a negative integer. From the discussion
after \eqref{eq:L-cont}, we see that
\[
  \lim_{N\to\infty}N^{-\alpha}L_m^\alpha(X)
\]
is only finite when $m\geq -\alpha$. We thus have to treat the terms
with $m < -\alpha$ separately. Expanding the polynomial in $X$, we
find
\begin{align}
  \sum_{m=0}^{-\alpha-1} L_m^\alpha(X)L_{N-m}^\beta(Y)
    &= \sum_{m=0}^{-\alpha-1}L_{N-m}^\beta(Y)
      \sum_{p=0}^m\frac{(-)^m}{p!}{-\alpha-p-1 \choose m-p}
      X^p\nonumber\\
    &= \sum_{p=0}^{-\alpha-1}\frac {X^p}{p!}
      \sum_{m=p}^{-\alpha-1}(-)^m{-\alpha-p-1 \choose m-p}L_{N-m}^\beta(Y)
      \nonumber\\
    &= \sum_{p=0}^{-\alpha-1}\frac {X^p}{p!}
      \findiff^{-\alpha-p-1} L_{N-p+\alpha+1}^\beta(Y).
    \label{eq:poly-to-findiff}
\end{align}
Here $\findiff$ is the forward difference, i.e. $\findiff f_m = f_{m+1}-f_m$,
and we have used that ${n \choose k} = (-)^k{-n+k-1 \choose k}$. 
It is easy to simplify the  difference operator acting on the
Laguerre polynomial. Induction
on the identity $L_m^{\alpha-1}(z)=L_m^\alpha(z)-L_{m-1}^\alpha(z)$ yields
\be
  \findiff^k L_m^{\alpha}(z) = L_{m+k}^{\alpha-k}(z).
  \label{eq:laguerre-fin-diff}
\ee
There is no further cancellation among the polynomials. The limit of 
\eqref{eq:poly-to-findiff} is therefore straightforward, and we find
\[
  \lim_{N\to\infty}N^{-\alpha-\beta-1}
      \sum_{m=0}^{-\alpha-1} L_m^\alpha(X)L_{N-m}^\beta(Y)
    = 2^{\alpha+\beta+1}y^{-\alpha-\beta-1}
      \sum_{j=0}^{-\alpha-1}\frac 1{j!}\left(\frac{-x^2}{2y}\right)^j
      J_{\alpha+\beta+j+1}(y).
\]
Note that these individual terms give contributions of the same order as the
rest of the sum.
With the difficult terms explicitly dealt with, the considerations leading
to \eqref{eq:g-sonine-sans-anoms}
goes through for the remaining terms, and we end up with the identity
\begin{multline}
  \frac 1 2 x^{-\alpha}y^{-\beta}\int_0^1\ud r\,
      r^{\alpha/2}(1-r)^{\beta/2}J_\alpha(\sqrt r x)J_\beta(\sqrt{1-r}y)
  = (x^2+y^2)^{-(\alpha+\beta+1)/2}J_{\alpha+\beta+1}(\sqrt{x^2+y^2})\\
  - y^{-\alpha-\beta-1}
      \sum_{j=0}^{-\alpha-1}\frac 1{j!}\left(\frac{-x^2}{2y}\right)^j
      J_{\alpha+\beta+j+1}(y)
  - x^{-\alpha-\beta-1}
      \sum_{j=0}^{-\beta-1}\frac 1{j!}\left(\frac{-y^2}{2x}\right)^j
      J_{\alpha+\beta+j+1}(x)
  \label{eq:g-sonine},
\end{multline}
which is valid both when $\Re \alpha > -1$ (with the sum understood as being zero), and
when $\alpha$ is a negative integer. The same applies to $\beta$. 
We see that the integral given by the LHS of \eqref{eq:g-sonine} is discontinuous 
at $\alpha=-1$ (approaching from the $\Re \alpha > -1$ half-plane), with the size
of the discontinuity given by the sum term on the RHS.
This is related to the disappearance of the branch cut of $J_\alpha(z)$ along the
negative real axis in the $z$ plane at $\alpha=-1$.

Taking $\alpha=0$ and $\beta=-1$, we obtain 
\[
  \frac y 2\int_0^1\ud r\,
      (1-r)^{-1/2}J_0(\sqrt r x)J_1(\sqrt{1-r}y)
    = -J_0(\sqrt{x^2+y^2})+J_0(x),
\]
which was previously obtained in the context of a random matrix calculation\cite{AD}.
This was extended to the series of identities with $\alpha=\beta+1$
a negative integer in ref. \onlinecite{AI}, but we have not found the general result
\eqref{eq:g-sonine} in the literature.

The example just given was special in that we could compute the sum of
the terms we had to separate out explicitly at finite-$N$.
In general, this will not be possible.
To illustrate how to handle this, let us consider the identity
(48.21.14 in ref. \onlinecite{Hansen})
\[
  \sum_{m=0}^N\frac{\left(\frac 1 2 [\alpha-\beta]\right)_m}{(\alpha+1)_m}
    L_m^\alpha(-X)L_{N-m}^\beta(X)
    = \frac{\left(\frac 1 2[\alpha+\beta]+1\right)_N}{(\alpha+1)_N}L_N^\alpha(X),
\]
where $(x)_m := x(x+1)\cdots(x+m-1)$ denote the rising factorial (or Pochhammer symbol).
For $\alpha,\beta\geq 0$ we can, mutatis mutandis, repeat the above analysis.
The only additional ingredient is that we need to use Stirling's approximation
for the factorials. The result is
\be
  \frac{2^\beta x^{-\beta}}{\Gamma\left(\frac 1 2[\alpha-\beta]\right)}
    \int_0^1\ud r\, r^{-\beta/2-1}(1-r)^{\beta/2}I_{\alpha}(\sqrt r x)
    J_\beta(\sqrt{1-r} x)
  = \frac 1 {\Gamma\left(\frac 1 2[\alpha+\beta]+1\right)}J_\alpha(x),
  \label{eq:IJ-sans-anom}
\ee
which appears to be a new identity.

As before, we can also consider the case where $\beta$ is a negative integer.
We again find that the terms with $N-m < -\beta$ are anomalous, but in
this case we will apply a more general technique to calculate the large-$N$
limit. By a rewriting similarly to what was done in \eqref{eq:poly-to-findiff},
we have
\[
  \sum_{m=N+\beta+1}^N\frac{\left(\frac 1 2 [\alpha-\beta]\right)_m}{(\alpha+1)_m}
    L_m^\alpha(-X)L_{N-m}^\beta(X) 
  = \sum_{p=0}^{-\beta-1}\frac{(-X)^p}{p!}\findiff^{-\beta-1-p}_m
    \left[\frac{\left(\frac 1 2 [\alpha-\beta]\right)_m}{(\alpha+1)_m}
    L_m^\alpha(-X)\right]_{m=N+\beta-1}.
\]
Lacking the analog of \eqref{eq:laguerre-fin-diff}, we need a different approach.
As shown in the appendix, we can, assuming 
some regularity of the finite-$N$ corrections, calculate the large-$N$ limit.
The prescription amounts to taking the limit of the expression
in the square bracket for general $m$, and then effectively 
substitute $\findiff_m\to N^{-1}\partial_r$.
Carrying this out, we have
\begin{multline*}
  \lim_{N\to\infty}N^{-\frac 1 2(\alpha+\beta)}
    \sum_{m=N+\beta+1}^N\frac{\left(\frac 1 2 [\alpha-\beta]\right)_m}{(\alpha+1)_m}
    L_m^\alpha(-X)L_{N-m}^\beta(X)\\
  = 2^\alpha\frac{\Gamma(\alpha+1)}{\Gamma\left(\frac 1 2[\alpha-\beta]\right)}
    x^{-\alpha}\sum_{p=0}^{-\beta-1}\frac 1 {p!}\left(\frac{-x^2}{4}\right)^p
    \left.\frac{\partial^{-\beta-1-p}}{\partial r^{-\beta-1-p}}\right|_{r=1}
    \left[r^{-\beta/2-1}I_\alpha(\sqrt r x)\right],
\end{multline*}
which lead us to generalize \eqref{eq:IJ-sans-anom} to
\begin{multline}
    2^\beta x^{-\beta}\int_0^1\ud r\, r^{-\beta/2-1}(1-r)^{\beta/2}I_{\alpha}(\sqrt r x)
    J_\beta(\sqrt{1-r} x)\\
    +\sum_{p=0}^{-\beta-1}\frac 1 {p!}\left(\frac{-x^2}{4}\right)^p
    \left.\frac{\partial^{-\beta-1-p}}{\partial r^{-\beta-1-p}}\right|_{r=1}
    \left(r^{-\beta/2-1}I_\alpha(\sqrt r x)\right)
  = \frac {\Gamma\left(\frac 1 2[\alpha-\beta]\right)}
   {\Gamma\left(\frac 1 2[\alpha+\beta]+1\right)}J_\alpha(x).
\end{multline}
Convergence of the integral requires $\Re[\alpha-\beta]>0$ and $\Re \beta > -1$,
or that $\beta$ is a negative integer. Let us give the identity for the first few
negative values of $\beta$ explicitly
\[
  -\frac x 2 \int_0^1\ud r\, r^{-1/2}(1-r)^{-1/2}I_{\alpha}(\sqrt r x)
    J_1(\sqrt{1-r} x) + I_\alpha(x) = \frac {\Gamma\left(\frac 1 2[\alpha+1]\right)}
   {\Gamma\left(\frac 1 2[\alpha-1]+1\right)}J_\alpha(x),
\]
\[
  \left(\frac x 2\right)^2\int_0^1\ud r\, (1-r)^{-1}I_{\alpha}(\sqrt r x)
    J_2(\sqrt{1-r} x)+\frac x 4 \left(I_{\alpha-1}(x)-xI_\alpha(x)+I_{\alpha+1}(x)\right)
    = \frac {\Gamma\left(\frac 1 2[\alpha+2]\right)}
   {\Gamma\left(\frac 1 2[\alpha-2]+1\right)}J_\alpha(x)
\]
and
\begin{multline*}
  -\left(\frac x 2\right)^3\int_0^1\ud r\, r^{1/2}(1-r)^{-3/2}I_{\alpha}(\sqrt r x)
    J_3(\sqrt{1-r} x)-\frac{x^3}8 I_{\alpha-1}(x)\\
    +\frac{8(\alpha^2-1)+4(\alpha+1)x^2+x^4}{32}
      I_\alpha(x) 
  = \frac {\Gamma\left(\frac 1 2[\alpha+3]\right)}
   {\Gamma\left(\frac 1 2[\alpha-3]+1\right)}J_\alpha(x).      
\end{multline*}
\section{Additional identities}
We briefly mention some other identities that we found interesting. 

Let us first consider the formula (eq. 48.3.13 of ref \onlinecite{Hansen})
\[
  \sum_{m=0}^N\frac{(-N)_m(\nu+1/2)_m}{m!(1/2-N)_m}
    L_{2m+2\nu}^{-2\nu}(2X) = \frac{(N+\nu)}{(1/2)_N(1/2)_\nu}
    \left[L_{N+\nu}^{-\nu}(X)\right]^2,
\]
with $\nu$ a non-negative integer. The large-$N$ limit leads to
\be
  \int_0^1\ud r\, [r(1-r)]^{-1/2}J_{2\nu}(2\sqrt r x) 
    = \pi\left[J_{\nu}(x)\right]^2,
  \label{eq:pi-ident}
\ee
where we have used that, since $\nu$ is an integer, $J_{-\nu}=(-)^\nu J_\nu$. 
If we allow $\nu$ to be a general complex number (with $\Re\nu > -1/2$ to ensure
convergence), we cannot deduce from the derivation that the identity
should continue to hold. Numeric calculations, however, strongly suggest the it remains
valid on this enlarged domain.

The identities we have considered until now have involved sums over the degree of
the polynomials, but one can also consider sums involving the order.
Let us take 
\[
  e^Y\sum_{m=0}^\infty\frac{(-)^m}{m!}Y^m L_N^{\alpha+m}(X)
    = L_N^\alpha(X+Y)
\]
as an example.
The asymptotic relation \eqref{eq:L-assymp} we have used so far is for fixed order,
 so it is
does not bound the error of replacing Laguerres with Bessel functions in the 
sum over $m$. If we nevertheless proceed with the replacement, we are lead to 
\be
  x^{-\alpha}\sum_{m=0}^\infty\frac{(-)^m}{m!}\left(\frac{y^2}{2x}\right)^m
  J_{\alpha+m}(x)
  =[x^2+y^2]^{-\alpha/2}J_\alpha(\sqrt{x^2+y^2}).
  \label{eq:order-sum-ident}
\ee
We have verified this is a valid identity to high precision numerically.

Lastly, let us remark that the same idea
applies when starting from an integral Laguerre identities. As an example, 
let us take (eq. 22.13.13 of ref. \onlinecite{AS})
\[
  \int_0^X\ud Y\, Y^\alpha(X-Y)^{\beta-1}L_N^\alpha(Y)
    = \Gamma(\beta)\frac{\Gamma(N+\alpha+1)}{\Gamma(N+\alpha+\beta+1)}X^{\alpha+\beta}
    L_N^{\alpha+\beta}(X),
\]
where convergence requires $\Re\alpha > -1$, $\Re\beta > -2$. Since the integration
domain is bounded, taking the large-$N$ limit presents no difficulty, and we find
\be
  \int_0^x\ud y\, y^{\alpha+1}(x^2-y^2)^{\beta-1}J_\alpha(y) 
    = 2^{\beta-1}\Gamma(\beta)x^{\alpha+\beta}J_{\alpha+\beta}(x).
\ee

It is interesting to note that \eqref{eq:g-sonine-sans-anoms} can also be obtained
from an integral identity. Indeed, starting from\footnote{Note that the related
  formula 2.19.14.3 of ref. \onlinecite{Prud} appears to be incorrect.} 
(2.19.14.1 in ref. \onlinecite{Prud})
\be
  \int_0^1\ud r\, r^\alpha (1-r)^\beta L_m^\alpha\left(\frac r {4N}\right)
    L_n^\beta\left(\frac{1-r}{4N}\right)
  = \frac{(m+n)!}{m!n!}B(\alpha+m+1,\beta+n+1)
    L_{m+n}^{\alpha+\beta+1}\left(\frac 1 {4N}\right),
  \label{eq:sonine-laguerre-int-ident}
\ee
and letting $N,n,m\to\infty$ such that $n/N\to x^2$ and $m/N\to y^2$, we easily
recover the previous result. Note, however, that \eqref{eq:sonine-laguerre-int-ident}
only converges for $\Re\alpha,\Re\beta > -1$, so it does not seem possible
to obtain the generalization \eqref{eq:g-sonine} this way.

\section{Conclusion}
We have illustrated a general method for obtaining Bessel identities through
a number of examples. The identities we have given as examples appear to be new. 

\begin{acknowledgments}
Poul Henrik Damgaard and Gernot Akemann are thanked for valuable discussions and
comments on the manuscript.
\end{acknowledgments}

\appendix*
\section{Finite differences in the large-$N$ limit}
We will give a brief argument that, when the large-$N$ limit is well-behaved, 
the finite difference operator $\Delta$ effectively turns into a differential,
in a way that will be made precise.
For that purpose, let us consider a family of functions $f_N(z) : \Cbb\to\Cbb$ 
for each integer $N>0$.
Assume that we have the asymptotic expansion
\[
  f_N\left(\frac z N\right) \sim F(z) + \sum_{p=1}^\infty N^{-p}E_p(z),\qquad N\to\infty,
\]
with $F,E_p$ analytic. Then we have\footnote{While this is not an
  asymptotic expansion in $N^{-1}$, it holds that if one retains the first $k$ terms
  of the $p$ sum, the remainder will be $O(N^{-k-1})$.}
($l$ being a fixed integer)
\begin{align*}
  f_{N+l}\left(\frac z N\right) &= f_{N+l}\left(\frac {(1+l/N)z}{N+l}\right)\\
    &\sim F([1+l/N]z) + \sum_{p=1}^\infty (N+l)^{-p}E_p([1+l/N]z)
\end{align*}
For any $k > 0$ Taylor expansion yields
\begin{align}
  f_{N+l}\left(\frac z N\right) &= \sum_{p=0}^k\frac{\partial_z^p F(z)}{p!}\left(\frac {lz} N\right)^p
      +\sum_{p=1}^k N^{-p}\sum_{j=0}^{k-p}\frac{E_p^j(z)}{p!}\left(\frac l N\right)^j+O(N^{-k-1})\nonumber\\
    &= \sum_{p=0}^k l^p \left[\frac{\partial_z^p F(z)}{p!}\left(\frac {z} N\right)^p
      +\sum_{j=1}^{k-p}\frac{E_j^p(z)}{j!}N^{-p-j}\right]+O(N^{-k-1}),
  \label{eq:f-taylor}
\end{align}
with
\[
  E_p^j(z) := 
    \left.\frac{\partial^j}{\partial r^j}\right|_{r=1}
    \left[r^{-p}E_p(rz)\right]_{r=1}.
\]
The crucial property of \eqref{eq:f-taylor} is that the leading (in $N^{-1}$) 
coefficient of $l^p$ is just a derivative of $F(z)$. Using that, for any
polynomial $P_l := \sum_p N^{-p} a_p l^p$, we have
\[
  \findiff^p P_l = p!N^{-p}a_p + O(N^{-p-1})
\]
we get the result 
\[
  \lim_{N\to\infty}N^{-p}\findiff^p f_N(z)
   = \left.\frac{\partial^j}{\partial r^j}\right|_{r=1}F(rz).
\]

%

\end{document}